\documentclass[fleqn]{article}

\usepackage{amsmath,latexsym,amssymb,cite}
\usepackage{graphicx}

\newcommand{\bls}[1]{\renewcommand{\baselinestretch}{#1}}

\def\noi{\noindent}

\newcommand{\Title}[1]{\noi {\Large\bf #1}\\[1ex]}

\def\Aunames#1{\noi{\large\bf #1}}
\def\auth#1{${}^{#1}$}
\def\Addresses#1{\medskip\noi \protect
	\begin{description}\itemsep -3pt {\it #1} \end{description}}
\def\addr#1#2{\item[${}^{#1}$]{\it #2}}

\newcommand{\Abstract}[1]{\vskip 2mm \begin{center}
        \parbox{16.4cm}{\small\noi #1} \end{center}\medskip}

\def\email#1#2{\footnotetext[#1]{e-mail: #2}\addtocounter{footnote}{1}}

\def\nq{\hspace*{-1em}}
\def\nqq{\hspace*{-2em}}
\def\nhq{\hspace*{-0.5em}}

\def\cm{\hspace*{1cm}}
\def\inch{\hspace*{1in}}

\def\Jl#1#2{#1 {\bf #2},\ }

\def\ApJ#1 {\Jl{Astroph. J.}{#1}}
\def\CQG#1 {\Jl{Class. Quantum Grav.}{#1}}
\def\DAN#1 {\Jl{Dokl. AN SSSR}{#1}}
\def\GC#1 {\Jl{Grav. Cosmol.}{#1}}
\def\GRG#1 {\Jl{Gen. Rel. Grav.}{#1}}
\def\JETF#1 {\Jl{Zh. Eksp. Teor. Fiz.}{#1}}
\def\JETP#1 {\Jl{Sov. Phys. JETP}{#1}}
\def\JHEP#1 {\Jl{JHEP}{#1}}
\def\JMP#1 {\Jl{J. Math. Phys.}{#1}}
\def\NPB#1 {\Jl{Nucl. Phys. B}{#1}}
\def\NP#1 {\Jl{Nucl. Phys.}{#1}}
\def\PLA#1 {\Jl{Phys. Lett. A}{#1}}
\def\PLB#1 {\Jl{Phys. Lett. B}{#1}}
\def\PRD#1 {\Jl{Phys. Rev. D}{#1}}
\def\PRL#1 {\Jl{Phys. Rev. Lett.}{#1}}

\def\al{&\nhq}
\def\lal{&&\nqq {}}
\def\eq{Eq.\,}
\def\eqs{Eqs.\,}
\def\beq{\begin{equation}}
\def\eeq{\end{equation}}
\def\bear{\begin{eqnarray}}
\def\bearr{\begin{eqnarray} \lal}
\def\ear{\end{eqnarray}}
\def\earn{\nonumber \end{eqnarray}}
\def\nn{\nonumber\\ {}}

\def\nnn{\nonumber\\ \lal }

\def\yy{\\[5pt] {}}
\def\yyy{\\[5pt] \lal }
\def\eql{\al =\al}

\def\tst{\textstyle}

\def\fract#1#2{{\tst\frac{#1}{#2}}}

\def\half{{\fract{1}{2}}}

\def\e{{\,\rm e}}
\def\d{\partial}

\def\diag{\mathop{\rm diag}\nolimits}

\def\const{{\rm const}}
\def\eps{\varepsilon}

\def\then{\ \Rightarrow\ }

\def\D{{\mathbb D}}
\def\M{{\mathbb M}}

\def\S{{\mathbb S}}

\def\mn{_{\mu\nu}}
\def\MN{^{\mu\nu}}
\def\mN{_\mu^\nu}

\def\tT{{\widetilde T}}

\def\eqn#1{\eq\eqref{#1}}
\def\rf{\eqref}
\def\vac{{}_{\rm vac}}
\def\pvac{p_\bot{}_{\rm vac}}

\def\GR{general relativity}

\def\EEs{Einstein equations}
\def\Scw{Schwarzschild}
\def\RN{Reissner-Nordstr\"om}
\def\KS{Kantowski-Sachs}
\def\BT{Birkhoff theorem}
\def\sph{spherically symmetric}

\def\cosco{cosmological constant}
\def\elmag{electromagnetic}

\def\asflat{asymptotically flat}

\bls{1.07}
\begin{document}
\twocolumn[

\Title{The Birkhoff theorem and string clouds}

\Aunames{K.A. Bronnikov,\auth{a,b,c,1} S.-W. Kim\auth{d} and M.V. Skvortsova\auth{b,2}}
   
\Addresses{\small
\addr a {VNIIMS, Ozyornaya ul. 46, Moscow 119361, Russia}
\addr b	{Peoples' Friendship University of Russia,
	ul. Miklukho-Maklaya 6, Moscow 117198, Russia}
\addr c	{National Research Nuclear University ``MEPhI''
	(Moscow Engineering Physics Institute), Moscow, Russia}
\addr d 	{Dept. of Science Education, Ewha Womans University, Seoul 03760, Korea}
}

\Abstract
    {We consider \sph\ space-times in GR under the unconventional assumptions 
     that the spherical radius $r$ is either a constant or has a null gradient in the 
     $(t,x)$ subspace orthogonal to the symmetry spheres (i.e., $(\d r)^2 = 0$). 
     It is shown that solutions to the Einstein equations with $r = \const$ contain 
     an extra (fourth) spatial or temporal Killing vector and thus satisfy the \BT\ under 
     an additional physically motivated condition that the lateral pressure is functionally 
     related to the energy density. This leads to solutions that directly generalize the 
     Bertotti-Robinson, Nariai and Plebanski-Hacyan solutions. Under similar conditions,
     solutions with $(\d r)^2 = 0$ but $r\ne\const$, supported by an anisotropic fluid,
     contain a null Killing vector, which again indicates a Birkhoff-like behavior. 
     Similar space-times supported by pure radiation (in particular, a massless radiative 
     scalar field) contain a null Killing vector without additional assumptions, which leads to
     one more extension of the \BT.
     Exact radial wave solutions have been found (i) with an anisotropic fluid and (ii) with a 
     gas of radially directed cosmic strings (or a ``string cloud'') combined with pure radiation.
     Furthermore, it is shown that a perfect fluid with isotropic pressure and a massive 
     or self-interacting scalar field cannot be sources of gravitational fields with a null but 
     nonzero gradient of $r$.      
    }

]  
\email 1 {kb20@yandex.ru}
\email 2 {milenas577@mail.ru} 

\section{Introduction}

  The original Birkhoff theorem [1--3] stated that in general relativity
  (GR) a spherically symmetric gravitational field in vacuum is necessarily 
  static and thus reduces to the Schwarzschild solution.
  
  It was later found that the full \Scw\ space-time contains a
  T-region with a special \KS\ type metric, and the theorem in fact tells us
  that a \sph\ vacuum field in a particular region is either static or
  homogeneous, i.e., the \EEs\ imply an additional space-time symmetry 
  not postulated from the beginning. Theorems of this kind make easier 
  the solution process in many physically important situations and provide 
  their better understanding.

  The \BT\ was later extended to different types of geometries (with spherical, 
  planar, pseudospherical, cylindrical symmetries and diverse dimensions) and 
  different material sources of gravity (the \cosco, linear and nonlinear \elmag\
  fields, scalar fields, gauge fields, perfect fluids, etc.),  see, e.g., 
  \cite{steph03, we95, schm} and references therein.

  In particular, in \cite{bk1, bk2} some general conditions were found under which 
  the field equations lead to independence of the metric from a spatial or temporal
  coordinate, hence the existence of an additional Killing vector field, absent in the
  original problem setting. The approach of \cite{bk1, bk2} was extended in 
  \cite{we95} to multidimensional \GR\ with diverse geometries and material 
  sources of gravity.

  More general geometric conditions for the emergence of an additional Killing 
  vectors, connected with the existence of a conformal Killing-Yano tensor, were
  recently found in \cite{cky1,cky2}.

  Analogs of the Birkhoff theorem have also been found in a variety of
  extensions of \GR, see, e.g., the recent papers \cite{ext1, ext2, ext3} and 
  references therein. Extensions of the theorem to quantum gravity models
  have also been considered \cite{fil}.

  In the present study we return to \sph\ space-times in \GR. Our goal is to
  make clear what happens if we cancel one of the conditions used in a simple
  proof of the theorem, namely, that the gradient of the spherical radius $r$ as 
  a function of the temporal ($x^0 = t$) and radial ($x^1 =x$) coordinates 
  should not be null (that is, $(\d r)^2 \equiv \d_a r \d^a r \ne 0$, $ a = 0,1$). 
  This question is of certain interest even despite the existence of more advanced 
  Birkhoff-like theorems (e.g., \cite{cky1,cky2,barnes73,bona88}) and though the possible 
  geometries have been generally classified in \cite{goenner-70, mcI-72}; meanwhile, 
  the possible material sources of such geometries have not been described.

  If $(\d r)^2 = 0$ (in particular, if $r = \const$), the \EEs\ seem to be of wave 
  nature, and it is tempting to try to obtain wave solutions indicating a non-Birkhoff 
  situation. We will consider a sufficiently general choice of material sources 
  of gravity and really obtain some simple exact solutions which seem to be new 
  to our knowledge. It turns out that many kinds of matter are incompatible with
  the condition $(\d r)^2 = 0$ (e.g., a perfect fluid and a minimally coupled 
  scalar field with a nonzero self-interaction potential), while others, in particular, 
  some kinds of anisotropic fluids, pure radiation and a massless scalar field with 
  a null gradient, combined with a ``gas'' or ``cloud'' of radially directed cosmic 
  strings, lead to geometries with a null Killing vector, indicating a behavior in the 
  spirit of the \BT. 
  
  The paper is organized as follows. Section 2 presents the basic equations,
  Section 3 briefly reproduces the well-known proof of the \BT\ and discusses
  its straightforward extensions. Section 4 is devoted to consequences of the
  condition $r = \const$, which is a special case of $(\d r)^2 = 0$, while 
  Section 5 deals with a null but nonzero gradient of $r(x,t)$. In the latter case,
  exact wave solutions are obtained with such sources of gravity as 
  an anisotropic fluid, a \cosco, an \elmag\ field, a string cloud, a massless 
  scalar field and pure radiation. Section 6 summarizes and discusses the results.

\section {Basic relations}

  The general \sph\ metric can be written in the form\footnote
      {Our conventions are: the metric signature $(+\ -\ -\ -)$; the curvature tensor
      $R^{\sigma}{}_{\mu\rho\nu} = \d_\nu\Gamma^{\sigma}_{\mu\rho}-\ldots,\
      R\mn = R^{\sigma}{}_{\mu\sigma\nu}$, so that the Ricci scalar
      $R > 0$ for de Sitter space-time and the matter-dominated
      cosmological epoch; the system of units $8\pi G = c = 1$.}
  (see, e..g., \cite{LL, BR})
\bearr              \label{ds}
     ds^2 = \e^{2\gamma}dt^2 - \e^{2\alpha} dx^2 - r^2 d\Omega^2,
\nnn       \inch
    d\Omega^2 = d\theta^2 + \sin^2 \theta\, d\varphi^2,
\ear
   where $\alpha,\ \gamma, \ r$ are functions of $x$ and $t$, and $r$
   is the spherical radius, i.e., the curvature radius of a coordinate sphere 
   $x =\const,\ t =\const$. In \rf{ds} there
   is a freedom of choosing a reference frame (RF) and specific coordinates
   in a given RF (a congruence of timelike world lines) by postulating certain
   relations between the functions $\alpha,\ \gamma,\ r$.

   The \EEs\ can be written in the form
\beq                                        \label{EE-R}
        R\mN  = - \tT\mN \equiv - (T\mN- \half\delta\mN T),
\eeq
  where $T\mN$ is the stress-energy tensor (SET) of matter, and
  $T \equiv T^\alpha_\alpha$ is its trace. In particular, we consider the 
  \cosco\ as a special kind of matter.

 The nontrivial components of the \EEs\ for the metric  \rf{ds} are
\begin{subequations} \label{EE}
\bear        \nq                                                \label{00}
       R_0^0 \eql  \e^{-2\gamma}
       [2\ddot{r}/r + \ddot{\alpha} 
         + \dot{\alpha}{}^2 -\dot{\gamma} (2\dot{r}/r+\dot{\alpha})]
\nnn \ \
       -\e^{-2\alpha} [\gamma'' + \gamma'(2r'/r + \gamma' -\alpha')]
       = - \tT^0_0,
\ear
\bear  \nq		                    					\label{11}
       R_1^1 \eql \e^{-2\gamma}[\ddot{\alpha}
          + \dot{\alpha}(2\dot{r}/r - \dot{\gamma} + \dot{\alpha})]
\nnn \ \
       -\e^{-2\alpha} [2r''/r + \gamma'' + \gamma'^2
              -\alpha' (2r'/r + \gamma')] = - \tT^1_1,
\nnn
\yy     \nq                     \label{22}
       R_2^2 \eql 1/r^2 + \e^{-2\gamma}
       [\ddot{r}/r +  (\dot{r}/r) (\dot{r}/r -\dot{\gamma} +\dot{\alpha})]
\nnn \nhq 
       - \e^{-2\alpha}
          [r''/r + (r'/r) (r'/r + \gamma' -\alpha')] = - \tT^2_2,
\yy         \nq                         \label{01}
       R_{01} \eql (2/r)  [\dot{r}{}' + \dot{\beta}\beta'
	                - \dot{\alpha}r' - \dot{r} \gamma'] = - \tT_{01},
\ear
\end{subequations}
  where dots and primes stand for $\d/\d t$ and $\d/\d x$, respectively.

  The spherical radius $r(x,t)$ may be considered as a scalar field in the
  2D subspace of our space-time with the coordinates $x^a = (x^0, x^1) 
  = (t, x)$. Its gradient $\d_a r$ may be spacelike, timelike or null. 

  We are going to explore what happens if we omit one of the conditions
  often used in formulations of the \BT, the condition that $\d_a r$ is not null. 
  On the contrary, we will assume that in the whole 4D space-time or its region,
\beq                        \label{r-null}
       (\d r)^2 \equiv \e^{-2\gamma}{\dot r}{}^2 - \e^{-2\alpha}r'{}^2 =0
\eeq
  Since any 2D Riemannian metric is conformally flat, the
  coordinates $x$ and $t$ can always be chosen so that
  $\gamma(x,t) \equiv \alpha(x,t)$, and then the condition \rf{r-null}
  reduces to $\dot{r} = \pm r'$. Without loss of generality let us choose
  the plus sign (otherwise we can simply re-denote $x\to -x$).
  With $\alpha=\gamma$ and $\dot r = r'$, the \EEs\ \rf{EE} are 
  substantially simplified:
\begin{subequations} \label{EE-a}  
\bear        \nq                                                    \label{00a}
       R_0^0 \eql  \e^{-2\alpha}(4 \alpha_{uv} + R_{01}) = - \tT^0_0,
\yy                         					\label{11a}
       R_1^1 \eql \e^{-2\alpha}(4 \alpha_{uv} - R_{01}) = - \tT^1_1,
\yy    						                   \label{22a}
       R_2^2 \eql 1/r^2  = -\tT^2_2,
\yy         					                      \label{01a}
       R_{01} \eql (2/r)  [r_{uu} - 2 r_u \alpha_u] = - \tT_{01},
\ear
\end{subequations}
  where we have introduced  the null coordinates
\beq
            u = t + x, \qquad v = t -x,                         \label{u,v}
\eeq
  and the subscripts $u$ and $v$ denote $\d/\d u$ and $\d/\d v$, respectively.
  Since $\dot r = r'$, we deal with $r = r(u)$.  The metric takes the form
\bear                        \label{ds-uv}
            ds^2  \eql  \e^{2\alpha(t,x)} (dt^2 - dx^2) - r^2\, d\Omega^2
\nn
	           \eql \e^{2\alpha(u,v)} du\, dv - r^2(u)\, d\Omega^2.
\ear
  (In what follows, we still prefer to write the \EEs\ and the SET components 
  referred to the coordinates $t$ and $x$ for physical convenience, to be able 
  to discuss reference frames (RFs), densities and pressures. The tensor 
  indices 0 and 1 will thus refer, as before, to $t$ and $x$.) 

\section {The \BT}

  Let us first, for completeness, formulate and prove the theorem
  in its simplest form, including into consideration vacuum and
  electrovacuum space-times with a \cosco.

\medskip\noi
  {\bf Theorem 1.} {\sl Consider the \EEs\ for the metric \rf{ds} with
  a \cosco\ $\Lambda$ and a Maxwell \elmag\ field $F\mn$. Suppose
  that the gradient of $r(x,t)$ is spacelike or timelike in a 
  space-time region $\D$. Then, in a certain coordinate system in $\D$ 
  all metric coefficients in \rf{ds} are $t$-independent (the 
  metric is static) or $x$-independent (the metric is homogeneous).}

\medskip\noi
  {\bf Proof.} The only nonzero components of the Maxwell tensor $F\mn$
  compatible with spherical symmetry are $F_{01}= - F_{10}$ (radial
  electric fields) and $F_{23}= - F_{32}$ (radial magnetic fields), and the
  Maxwell equations without sources imply $F_{01}F^{10}= q_e^2/r^4$
  and $F_{23} F^{23} = q_m^2/r^4$ in the general metric \rf{ds},
  the constants $q_e$ and $q_m$ being the electric and magnetic
  charges, respectively. So the SET of the Maxwell field is
\beq                                						\label{SET-F}
      T\mN[F] = \tT\mN [F] = \frac{q^2}{r^4} \diag(1,\ 1,\ -1,\ -1),
\eeq
  where $q^2 = q_e^2 + q_m^2$. The contribution of the \cosco\ into the
  total SET $T\mN$ is $\Lambda\delta\mN$ (hence $-\Lambda\delta\mN$
  to $\tT\mN$) in any coordinate system. The total SET component
  $\tT_{01}$, corresponding to a radial energy flow, is equal to zero.

  Suppose that the vector $\d_a r$ is spacelike. Then $r$ can be chosen as 
  a radial coordinate, that is, $r = x$. Equation \rf{01} then gives 
  $r'\dot\alpha =0$, and since $r'=1$, we have $\alpha = \alpha(x)$.
  Thus only $\gamma$ may depend on $t$. However, \eq \rf{22}
  expresses $\gamma'$ via functions of $x$ only, hence
  $\gamma =\gamma_1(x) + \gamma_2(t)$. Lastly, $\gamma_2$ can 
  be turned to zero by a coordinate transformation $t \mapsto {\tilde t} (t)$. 
  Thus the metric is static.

  If the vector $\d_a r$ is timelike, quite a similar reasoning shows that
  in the coordinate system where $r = t$ the metric is $x$-independent,
  in other words, the space-time is homogeneous and belongs to the class
  of \KS\ cosmological models. This completes the proof. $\Box$

\medskip
  As is well known, both static and homogeneous solutions to the \EEs\ are
  then unified in the \RN-de Sitter metric
\bearr                          \label{RNdS}
    ds^2= A(r)dt^2 - dr^2/A(r) - r^2 d\Omega^2,
\nnn
    \cm A(r) = 1 - \frac{2m}r + \frac{q^2}{r^2} - \frac{\Lambda r^2}{3},
\ear
  whose special cases are the \Scw\ ($\Lambda=q=0$), \RN\ ($\Lambda=0$)
  and (anti-)de Sitter ($m=q=0$) metrics. Regions with $A(r) > 0$ are static, 
  those with $A(r) < 0$ are homogeneous \KS\ regions, and regular zeros of
  $A(r)$ correspond to Killing horizons whose number and nature depend on
  the values of the three parameters $m$, $q$, and $\Lambda$.

  A more general geometric formulation of the theorem is \cite{barnes73,bona88}:

\medskip\noi
  {\bf Theorem 2.}  {\sl Metrics with a group G3 of isometries on non-null 
  orbits V2 and with Ricci tensors of Segre types [(11)(1,1)] and [(111,1)] 
  admit a group G4, provided that $Y_{,a} \ne 0$.}

\medskip
  Some comments are in order. First, G3 with non-null orbits on V2 describe
  spherical, plane and pseudospherical symmetries, and $Y_{,a}$ corresponds
  in our notations to $\d_a r$. As already said, we here focus on spherical symmetry.

  Second, any Ricci tensor types coincide with types of the SET $\tT\mN$
  due to \eqs \rf{EE-R}. Accordingly,  type [(111,1)] means a \cosco, while the 
  more general type [(11)(1,1)] corresponds to any kind of matter with $T^0_0 
  = T^1_1$. This condition defines a {\it generalized notion of vacuum\/} as a kind
  of matter for which there is no unique comoving RF: instead, all RFs moving in a 
  certain direction, say, $x^1$ (the radial one in our case) are comoving 
  (the flux $T^1_0$ is zero) \cite{dym92, kb-dym03}. We will call such a kind 
  of matter {\it Dymnikova's vacuum\/} or, shorter, {\it D-vacuum}. Special cases of 
  D-vacuum are not only a \cosco\ and a Maxwell radial \elmag\ field but also, e.g., 
  nonlinear \elmag\ fields with Lagrangians of the form  $-L(F)$, $F = F\mn F\MN$
  (see, e.g., \cite{NED1, NED2, kb-NED}). In \cite{dym92} this kind of SET was
  argued to follow from the properties of quantum fields in curved space-time.

  Third, the \BT\ can be extended to some SETs not mentioned in Theorems 1 
  and 2 (e.g, perfect fluids, scalar fields, etc.) and symmetries like the cylindrical 
  one \cite{bk1, bk2}, though under additional conditions that prevent the emergence 
  of waves. Some cases of interest are listed in \cite{bk1, bk2}; such a result was 
  obtained there by finding more general conditions under which, according to the 
  \EEs, $\gamma'$ is $t$-independent or $\dot\alpha$ is $x$-independent 
  (see the above proof of Theorem 1).

  Fourth, the conditions of Theorem 2 exclude the assumption $r= \const$ 
  (to be discussed in Section 4) but admit a null gradient $\d_a r \ne 0$ 
  (to be discussed in Section 5); evidently, in the latter case the extra (fourth) Killing
  vector, which exists according to the theorem, should be null. We shall see that 
  such a null Killing vector also appears in solutions with a number of SETs 
  other than those mentioned in Theorem 2.

\section{Systems with $r = r_0 = \const$}

  The condition $r = \const$ looks somewhat unusual, but the corresponding 
  solutions are rather much discussed in the literature; examples of their possible
  application to certain regions of more plausible space-times are long wormhole 
  throats \cite{long} and ``horned particles'' \cite{horn, oz-09}.

  If $r = \const$, we have $R_{01} = 0$ and $R^0_0 = R^1_1$   (see \eqs 
  \rf{EE-a}), hence $T^0_0 = T^1_1$, i.e., only matter of D-vacuum 
  type is admitted. The most general SET of such matter reads
\beq
        T\mN = \diag (\rho, \rho, -p_\bot, -p_\bot),       \label{T-vac}
\eeq
  where $\rho$ is the density and $p_\bot$ the lateral pressure. For
  $\tT\mN$ it then follows
\beq                                                    			\label{tT-vac}
        \tT\mN = \diag (p_\bot, p_\bot, -\rho, -\rho).
\eeq 
  Evidently, this ``vacuum'' can consist of a few components, for example, 
  a \cosco\ and a Maxwell field as mentioned above, and anything 
  else with the same structure of the SET.

  Now, from \eqn{22a} it follows that 
\beq                                                             \label{rho0}
            \rho = 1/r_0^2 = \const. 
\eeq
  The conservation law $\nabla_\nu T\mN =0$ then holds automatically, 
  leaving $p_\bot$ quite an arbitrary function of $u$ and $v$. Furthermore, 
  \eqn{00a} has the form of a nonlinear wave equation for $\alpha(u,v)$:
\beq                                        \label{wave-eq}
    	4\alpha_{uv} =  -p_\bot \e^{2\alpha}.
\eeq
  With an arbitrary $p_\bot$, $\alpha(u,v)$ is also arbitrary, and no extra
  symmetry is observed. So the case $r = \const$ is correctly excluded from
  Theorem 2 (as remarked in \cite{bona88}). However, it is physically
  reasonable to suppose that $p_\bot$ is connected with $\rho$ by a kind of 
  equation of state, then $p_\bot = \const$, and \eqn{wave-eq} is a Liouville
  equation whose solutions are well known: according to \cite{PZ}, the 
  general solution for $p_\bot \ne 0$ is 
\begin{subequations} \label{sol0}
\bearr              				\label{gen}
        2\alpha(u,v) = f(u) + g(v)
\nnn\
       	 - 2 \ln\left|k\int \e^{f(u)}du - \frac{p_\bot}{4k}\int \e^{g(v)} dv\right|
\ear
  with arbitrary functions $f(u)$ and $g(v)$ and an arbitrary constant $k$.
  There are, in addition, five special solutions \cite{PZ}:
\bearr              \label{spec1}
           \e^{2\alpha} = - \frac{4ab}{p_\bot  S^2(z)},\qquad  z = au + bv,
\yyy \nq
      \mbox{with four variants of $S (z)$}: 
\nnn
           S(z) = \{z,\ \cos z,\  -\cosh z,\ \sinh z\},  \ \  {\rm and\ also} 
\nnn            \label{spec5}
    	 \e^{2\alpha} = - \frac{4C}{p_\bot (uv - C)^2},
\ear
\end{subequations}
  where $a,\ b,\ C$ are arbitrary nonzero constants. Note that $p_\bot$ can have
  any sign: for example, if matter in question consists of a \cosco\ and a Maxwell
  field, then
\beq           							\label{SET1}
	   \rho = \Lambda + q^2/r^4, \quad \ p_\bot  = - \Lambda + q^2/r^4.
\eeq
  In the case $p_\bot =0$, \eqn {wave-eq} simply gives $2\alpha(u,v) = f(u) + g(v)$,   
  and then the substitution $\e^{f(u)}du = dU$, $\e^{g(v)}dv = dV$ reduces the
  metric \rf{ds-uv} to the simplest form
\beq 								\label{ds0}
 	ds^2 = dU\,dV - r_0^2 d\Omega^2 =  dT^2 - dX^2 - r_0^2 d\Omega^2,
\eeq
  where $2T = U+V$ and $2X = U-V$, i.e., the geometry is simply
  $\M^2 \times \S^2$, where $\M^2$ stands for 2D Minkowski space. This solution
  not only corresponds to the special case $ \Lambda = q^2/r^4$ of \rf{SET1},
  but also to a kind of matter with a SET having the structure 
\beq 							\label{string}
          T\mN = \diag (\rho,\ \rho,\ 0,\ 0)
\eeq
   which can be ascribed to a distribution of cosmic strings aligned to the direction
   $x= x^1$, a ``string cloud'' for brevity (see, e.g., \cite{mh15} and references therein).

   In the pure vacuum case with $T\mN =0$, due to \rf{rho0}, there is no solution,
   i.e., for pure vacuum the condition $\d_a r \ne 0$ in Theorem 2 is unnecessary 
   (as was mentioned, e.g., in \cite{HEll}).

  Returning to the solutions \rf{sol0}, we notice that the same substitution
  $dU = \e^{f(u)} du$, $dV = \e^{g(v)} dv$ applied to the solution 
  \rf{gen} and a change in the notations $U\mapsto u,\ V\mapsto v$ result in
\beq                            \label{UV-gen}
    \e^{2\alpha} du\,dv = du\,dv \Big(ku - \frac{p_\bot}{4k}v \Big)^{-2}.
\eeq
  It coincides with the special solution \rf{spec1} with $S(z)=z$ and
  $a/b = -p_\bot/(4k^2)$. A curious observation is that a general solution 
  from the viewpoint of differential equations reduces to a special one from the
  viewpoint of space-time geometry.

  Furthermore, in all cases of  \rf{spec1} a further substitution 
  $au \mapsto u$ and $bv \mapsto v$ (or $bv \mapsto - v$ since $a$
  and $b$ can be positive or negative), leads to $au + bv \mapsto u \pm v$.
  Returning to $t$ and $x$ according to \eqn{u,v}, we see that in the new, 
  thus obtained coordinates the function $\alpha$ depends either on $t$ 
  only or on $x$ only.

  In \rf{spec5}, assuming $u > 0,\ v > 0$, we substitute
\beq
          u = \sqrt{C+y} \e^{-z/2}, \quad\   v = \sqrt{C+y} \e^{z/2},
\eeq
  and obtain
\beq                                           \label{sol-5a}
       \e^{2\alpha}du\,dv = \frac{C}{p_\bot y^2}
		     \biggl(-\frac{dy^2}{C+y} + dz^2 \biggr).
\eeq
  Thus the metric depends on the single coordinate $y$ which can be spatial
  or temporal depending on the sign of the factor $C/p_\bot$. If $v$ or/and 
  $u$ are negative, the same substitution will work with $-u$ or/and $-v$ 
  instead of $u$ and $v$, respectively.

  Thus in all such cases the resulting 2D metric only depends on one of the
  coordinates, spatial or temporal, indicating the existence of the fourth, temporal
  or spatial Killing vector. In other words, the \BT\ holds under the assumption
  $r=\const$ under the additional (physically motivated) assumption 
  $p_\bot = \const$, but one need not assume the Ricci tensor structure as
  in Theorem 2 since this structure now follows from the \EEs.

  As to the specific form of static and homogeneous (\KS) solutions 
  for $\alpha(x)$ or $\alpha(t)$, it is more convenient, instead of transforming 
  different cases of \eqs\rf{sol0}, to find them directly from \eqn{00a} 
  (coinciding with \rf{11a}). We thus have the Liouville equation 
\beq               							\label{Liou}
	\alpha'' = p_\bot \e^{2\alpha}, \ \ {\rm or}\ \
		\ddot{\alpha} = - p_\bot \e^{2\alpha}.
\eeq
  with $p_\bot = \const \ne 0$. Its solutions are well known and lead to
  straightforward generalizations of the Bertotti-Robinson \cite{bert, rob},
  Nariai \cite{nari} and Pleba\'nski-Hacyan \cite{PleHa} space-times
  to more general cases of D-vacuum. Different cases and properties 
  of these solutions are quite well described in the literature (see, e.g., 
  \cite{griff} and references therein) and are beyond the scope of this paper. 
  
\section{Systems with null nonzero $\d_a r$}
\subsection{Comoving matter}

  Let us try to reveal which sources of gravity are compatible with variable $r$
  satisfying $(\d r)^2=0$. We will first try to simplify the problem assuming 
  that the corresponding coordinate system $(t,x)$ belongs to the comoving 
  reference frame of matter under consideration. It is manifestly the case for 
  any kind of D-vacuum, including the \cosco\ and the radial Maxwell 
  fields, but it is an additional assumption for other kinds of matter.\footnote
	{The form \rf{ds-uv} of the metric selects a certain class of RFs,
	connected by Lorentz boosts in the $(x,t)$ Minkowski plane and their 
	conformal extensions corresponding to substitutions $u=u(U),\ v=v(V)$.}
  As mentioned before, we have $r = r(u)$.

  Comoving means that the energy flow is zero, hence due to the \EEs\ 
  $R_{01} = 0$, and by \eqn{01a}   
\beq
         r_{uu} = 2\alpha_u r_u,                                     \label{01b}
\eeq
   The remaining \EEs\ \rf{EE} lead to the following relations:
\bearr
    \tT^0_0 = \tT^1_1 = 4\e^{-2\alpha} \alpha_{uv},       \label{001b}
\yyy
        \tT^2_2 = - e^{-2\beta} = - 1/r^2(u).             \label{22b}
\ear

  Since by assumption $r \ne \const$, according to \rf{01b} $\alpha_u$ 
  is a function of $u$ only, and
\beq             \label{alpha}
    	\e^\alpha = |r_u|\cdot A_1(v).
\eeq
  Choosing properly the coordinate $v$, we achieve $A_1 \equiv 1$, while 
  a transformation $u  =  u(U), \ U\mapsto u$ allows us to put $r = u$, hence
  $\e^\alpha = 1$. Thus the metric is completely known and has the simple form
\bear              							 \label{ds-w}
        ds^2 \eql    du dv - u^2 d\Omega^2
\nn
               \eql dt^2 - dx^2 -(t+x)^2 d\Omega^2.
\ear

   Furthermore, by \rf{001b} $R^0_0=R^1_1 =0$, and the total SET 
   necessarily has the form \rf{string}, where, according to \rf{22b},
\beq
 	\rho = 1/u^2.				\label{rho-u}
\eeq   
  We see that the only admissible kind of matter able to support
  this gravitational field has the SET \rf{string} and is interpreted as a 
  cloud of radially aligned cosmic strings.\footnote
	{As mentioned above, rather a general form of D-vacuum is represented 	by nonlinear 
	\elmag\ fields with Lagrangians of the form $-L(F)$, $F\equiv F\mn F\MN$. It is of 
         interest to know which $L(F)$ corresponds to a ``stringy'' D-vacuum with $\pvac =0$.
         Assuming that there are both electric and magnetic fields with charges $q_e$ and
	$q_m$, respectively, it can be shown that the condition $\pvac = -T^2_2=0$ leads to  
         the following differential equation for $L(F)$:
\[
           L (q_m^2\, L_F^2 - q_e^2 ) = 2 q_m^2\, F L_F^3,  
\]     
         where $L_F \equiv dL/dF$. It is hard to solve in general, but a simple solution 
         is obtained in the case $q_e =0$ (a pure magnetic field): $L = \sqrt{F/F_0}$, 
         $F_0 = \const$, a kind of Lagrangian used, e;g., in \cite{horn}.
         If, on the contrary, $q_m =0$, 	we obtain the condition $L=0$, which can only hold 
	for a nonzero electric field if this field is constant, which in turn requires $r = \const$. 
 	}

\subsection{A noncomoving fluid}

  Now let us assume that there is matter compatible with $(\d r)^2 =0$
  and having a nonzero SET component $T_{01}$. We will consider a few 
  kinds of such matter. (Recall that we consider the SET components 
  corresponding to the coordinates $t$ and $x$, for example,
  $T_{01}\equiv T_{tx}$.) Since $r = r(u) \ne \const$, we again choose
  the null coordinate $u$ so that $r \equiv u$ (at least in a certain range of $u$).

  In all cases we will admit that, in addition to the sort of matter under
  consideration, the sources of gravity include some sort of D-vacuum 
  with the SET (\ref{T-vac}), or even a combination of different
  D-vacua not interacting with matter or with each other.  
  As already mentioned, all of them are comoving to any 
  radially moving RF and do not contribute to $T_{01}$ and $T^0_0 - T^1_1$. 
  Their density and lateral pressure will be denoted $\rho\vac$
  and $\pvac$, respectively, to distinguish them from the noncomoving matter.

  Let us begin with an (in general) anisotropic fluid. The SET has the form
  \cite{Let87}
\beq                         						\label{SET-an}
	T\mN = (\rho + p_\bot) u_\mu u^\nu - \delta\mN p_\bot 
			+ (p_r - p_\bot)\chi_\mu \chi^\nu,
\eeq
  where $\rho,\ p,\ u^\mu$ are the density, pressure and (timelike) 4-velocity,
  respectively, and $\chi^\mu$ is a spacelike unit vector in the velocity direction.
  By symmetry of the problem, $u^2 = u^3 =0$, and the normalization 
  condition $u^\alpha u_\alpha =1$ can be written in the form 
\beq                                                      \label{norm}
               u_0 u^0 + u_1 u^1 =  \e^{-2\alpha}(u_0^2 - u_1^2)  =1.
\eeq
  Also, $\chi^\nu = (0, \e^{-\alpha}, 0, 0)$.
  Now, independent combinations of the \EEs\ \rf{EE-a} can be written as follows:
\begin{subequations}        \label{EE-f}
\bear                                                  \label{01f}
	4 \frac{\alpha_u}{u} \eql  (\rho + p_\bot) u_0 u_1,
\yy                                                       \label{0-1f}
	8 \frac{\alpha_u}{u} \eql (\rho + p_\bot)(u_0^2 + u_1^2) + \e^{2\alpha}(p_r-p_\bot),
\yy                                                        \label{0+1f}
	4 \alpha_{uv} \eql   -\e^{2\alpha}(p_\bot + \pvac),
\yy  					 \label{22f}
	\frac{2}{u^2} \eql  \rho - p_r + 2\rho\vac,
\ear	
\end{subequations}   
  where \eqn{01f} is the ${1 \choose 0}$ component, 
  \eqs \rf{0-1f} and \rf{0+1f} are a difference and a sum of ${0 \choose 0}$ and
  ${1 \choose 1}$ components, and \rf{22f} is the ${2 \choose 2}$ component 
  of  \rf{EE-a}. Subtracting doubled (\ref{01f}) from (\ref{0-1f}),  we obtain
\beq
	 \e^{2\alpha}(p_\bot - p_r) = (\rho + p_\bot) (u_0 - u_1)^2.       \label{u01}
\eeq
  The opportunities $\rho + p_\bot =0$ or $p_\bot = p_r$ should be rejected   
  since they lead either to the SET of a \cosco\ (which is comoving 
  to any RF) or to $u_0 = u_1$, making $u^\mu$ a null vector. In both cases
  it is not a noncomoving fluid.    

  Thus, in particular, {\it a Pascal perfect fluid with $p_\bot = p_r$ cannot be a 
  source of the metric under consideration.} This conclusion is independent of a 
  possible presence of any kind of D-vacuum since the latter does not 
  contribute to \eqs (\ref{01f}) and (\ref{0-1f}).

  There is a considerable freedom in choosing the form of matter variables which 
  can lead to metrics with $\alpha = \alpha (u,v)$. Let us, however, show that 
  under some natural assumptions there is no $v$ dependence, and we again 
  obtain the fourth ($v$-directed) Killing field, i.e., a Birkhoff-like situation.

  Suppose that (i) there are equations of state $p_r = p_r (\rho)$ and  
  $p_\bot = p_\bot (\rho)$ and (ii) $\rho\vac$ and $\pvac$ are functions of 
  $r$, hence of $u$. (Note that it is true for a \cosco\ and an \elmag\ field.) 
  From \rf{22f} it is then clear that $\rho$, $p_r$ and $p_\bot$ are functions 
  of $u$, and from \rf{u01} with \rf{norm} it follows that 
  $\e^{-\alpha} u_0$ and $\e^{-\alpha} u_1$ are functions of $u$. Further, from \rf{01f} 
  we find that $(\e^{-2\alpha})_u$ is a function of $u$, hence
\bearr 								\label {A1}
	(\e^{-2\alpha})_{uv} = 0 \ \then \ \alpha_{uv} = 2 \alpha_u\alpha_v,
\yyy									\label {A2}
        \e^{-2\alpha} = U(u) + V(v), \qquad U,V = {\rm arbitrary}.
\ear
  Substituting \rf{A1} into \eqn{0+1f}, we obtain
\beq 									\label {0+1ff}
            4 U_u \alpha_v = p_\bot + \pvac,
\eeq
  If $U(u) =\const$, we have $\alpha=\alpha(v)$ which is converted 
  to $\alpha \equiv 0$ by rescaling of $v$, and then by \rf{0+1ff}, 
  $p_\bot + \pvac =0$. If, on the contrary, $U(u) \ne \const$, then 
  we can equate two different expressions for $\alpha_v$ that follow 
  from \rf{A2} and \rf{0+1ff}, to obtain
\beq          							\label{Vf}
	V_v = -\frac {p_\bot + \pvac}{2 U_u} (U+V).
\eeq 
  Since $U_u \ne 0$, applying $\d_u$ to both parts of 
  this equality, we see that it can hold only with $p_\bot + \pvac =0$
  and $V = \const$.

  Thus under the assumptions made we find that the field 
  equations inevitably lead to (i) $\alpha = \alpha(u)$ and 
  (ii) $p_\bot + \pvac =0$.   

  To verify that such solutions do really exist, let us give a simple 
  example, assuming that there is no ``vacuum'', $p_\bot =0$, and 
  $p_r = -w \rho$ with $w = \const \in (0, 1)$ (the inequality $p_r < 0$ 
  follows from \rf{u01} with $p_\bot =0$). Then from \eqs \rf{22f},
  \rf{u01} and \rf{norm} we obtain
\beq                                                                  \label{sol-f}
          \rho = \frac{2}{(1{+}w)u^2}, \quad 
          u_0 = \e^\alpha \frac {1{+}w}{2\sqrt{w}}, \quad
          u_1 = \e^\alpha \frac {1{-} w}{2\sqrt{w}}.
\eeq
  It remains to find $\e^\alpha (u)$, which can be done using \rf{01f}
  with $\rho$ substituted from \rf{sol-f}. Integrating it, we obtain
\beq
         \e^{-2\alpha} = -\frac{1-w}{4w} \ln \frac{u}{u_*}, 
		\quad\ u_* = \const.
\eeq
  It is of interest that at $u= u_*$ there is an apparent singularity, where
  $\e^\alpha \to \infty$; however, one can verify that all curvature invariants
  remain finite.

  Other solutions with anisotropic fluids, for example, those with a \cosco\ 
  or a string cloud added, can also be found. However,  due to 
  \eqn{0+1f}, in all such cases the total lateral pressure is equal to zero.  

 \subsection{Pure radiation} 

  We have shown that there are solutions with a noncomoving anisotropic
  fluid. Let us now consider such matter that has no comoving RF at all, 
  namely, pure radiation whose SET has the form $T\mN = \Phi(u,v) k_\mu k^\nu$, 
  where $\Phi$ is a scalar function and $a_\mu$ is a radial null vector, which, 
  since there are two possible directions, can be chosen in one of the two forms
\bear 
	k^\mu_{(\pm)} \eql (\e^{-\alpha}, \pm \e^{-\alpha}, 0, 0),
\nn
	k_{\mu (\pm)} \eql (\e^{\alpha}, \mp \e^{\alpha}, 0, 0),
\ear
  where the upper sign corresponds to a flow directed to larger values of $x$ 
  and the lower sign to smaller ones. A general situation is the existence 
  of two opposite radiation flows, with the full SET 
\beq                                            \label{SET-rad}
	T\mN = \Psi(u,v) k_{\mu(+)} k^\nu_{(+)} + \Phi (u,v) k_{\mu(-)} k^\nu_{(-)},
\eeq
  where $\Phi_\pm$ are scalar functions. In the metric \rf{ds-uv}, this SET has the 
  following nonzero components:
\bearr       								\label{SET-r}
	(T_a^b) = \begin{pmatrix} T^0_0 & T^0_1\\
					    T^1_0 & T^1_1    \end{pmatrix}
          = \begin{pmatrix}    \Phi + \Psi  &  \Phi  - \Psi \\
					       -\Phi + \Psi   &  - \Phi - \Psi  \end{pmatrix},
\nnn \cm
		  a, b = 0, 1 \equiv t, x,   
\ear
  while all other $T\mN$ are zero. Note that the trace $T_\mu^\mu$ is zero, therefore,\
  (see \rf{EE-R}) $\tT\mN = T\mN$. With this SET, \eq (\ref{01a}) and the difference 
  of  (\ref{00a}) and (\ref{11a}) lead to the following relations:
\bear                                                        \label{01r}
	4\alpha_u/u  \eql  \e^{2\alpha} (\Phi  - \Psi),
\nn
	4\alpha_u/u  \eql  \e^{2\alpha} (\Phi  + \Psi),
\ear
  whence it immediately follows $\Psi = 0$. Thus {\it the metric \rf{ds-uv} is only 
  compatible with a radiation flow directed to decreasing values of $x$}, which is in fact
  the propagation direction of the wave of the quantity $r$ since $r = u = x+t$. 

  In what follows we consider $T\mN = \Phi (u,v) k_\mu k^\nu$ with  
  $k^\mu = k^\mu_{(-)}$. From \rf{01r} we now have 
\beq     								\label{01rr}
	4\alpha_u/u =  \e^{2\alpha} \Phi (u,v),
\eeq
  and the other \EEs, (\ref{22a}) and a sum of (\ref{00a}) and (\ref{11a}) give
\bearr                                                      \label{22r}
	1/u^2 =  \rho\vac,
\yyy                                                        \label{0+1r}
	4 \e^{-2\alpha} \alpha_{uv} =  - \pvac.
\ear
  Equation (\ref{22r}) shows, in particular, that the inclusion of
  $\rho\vac$ is quite necessary for the existence of a solution. Even more than 
  that: not each kind of ``vacuum'' is suitable, for example, a combination of
  a \cosco\ and a Maxwell field gives $\rho\vac = \Lambda + q^2/u^4$,
  and \eqn{22r} then leads to  $u = \const$ which is meaningless. On the
  contrary, a string cloud \rf{string} is suitable, with it \eqn{22r} simply
  expresses $\rho\vac= \rho_{\rm string}$ in terms of $u$. 

  Lastly, the conservation equation $\nabla_\nu T\mN$ applied to the SET
  under consideration gives
\beq                                     					\label{Phi}
	\Phi_v  = -2\Phi \alpha_v =0 \ \then \ \Phi = F(u) \e^{-2\alpha}
\eeq
  with an arbitrary function $F(u)$. Its comparison with \rf{01rr}
  gives $\alpha_u = u F(u)/4$, a function of $u$ only; its further 
  integration leads to $\alpha$ with an additive arbitrary function of $v$
  which can be, as usual, absorbed by rescaling of the coordinate $v$, so 
  that $\alpha = \alpha (u)$ without loss of generality.   

  This completes the solution. We have a single arbitrary function
  $\alpha(u)$, the radiation flux density $\Phi$ is then obtained from \rf{Phi}
  with $F = 4\alpha_u/u$. Furthermore, from \rf{0+1r} and \rf{22r} it follows 
  $\pvac =0$ and $\rho\vac = 1/u^2$, so that again the only admissible kind of 
  D-vacuum is a string cloud. 

  We also conclude that pure radiation as a source of gravity with the metric 
  \rf{ds-uv} {\it automatically} leads to a Birkhoff situation with a null 
  ($v$-directed) additional Killing vector.    

\subsection{Scalar fields} 
  
 Consider a scalar field $\phi (x,t) = \phi(u,v)$ with the Lagrangian
\beq                                                       \label{L-s}
	L_s = \eps g\MN \d_\mu \phi \d_\nu\phi - 2V(\phi).
\eeq
  where $\eps=\pm 1$, $\eps=1$ corresponds to a normal, canonical scalar
  field, $\eps=-1$ to a phantom one, and $V(\phi)$ is its self-interaction
  potential. The modified SET $\tT\mN$, entering into the right-hand side
  of \eqs (\ref{EE}), has the form
\beq                                                      \label{SET-s}
	\tT\mN = \eps \d_\mu\phi \d^\nu\phi - \delta\mN V(\phi).
\eeq
  The scalar field equation following from (\ref{L-s}) is
\beq                                                       \label{eq-sc}
	\eps \Box \phi + dV/d\phi =0,
\eeq
  where $\Box = \nabla^\mu \nabla_\mu$ is the d'Alembert operator. In the
  metric (\ref{ds-uv}) it takes the form
\beq                                                       \label{eq-s}
	4 \eps\e^{-2\alpha}\biggl(\phi_{uv}
  			+ \frac{1}{u}\phi_v\biggr) + \frac{dV}{d\phi} =0.
\eeq
 
  Now, \eqn{01a} and the difference of (\ref{00a}) and (\ref{11a}) take 
  a form similar to \rf{01r} 
\bear                                                     \label{01s}
	4\alpha_u/u \eql \eps(\phi_u^2 - \phi_v^2),
\nn                                                       
	4\alpha_u/u \eql \eps(\phi_u^2 + \phi_v^2),
\ear
  whence it follows $\phi_v =0$, so that $\phi = \phi(u)$.
  Then $\Box \phi =0$ (see (\ref{eq-s}) and (\ref{eq-sc}), and consequently
  $dV/d\phi =0$. Thus {\it the scalar field can only be massless and
  $u$-dependent}. (A possible constant potential is simply an addition to the
  \cosco.)

  A further substitution to \eqs (\ref{22a}) and a sum of (\ref{00a}) and
  (\ref{11a}) (the latter takes the form \rf{0+1r}) lead to expressions 
  for $\rho\vac$ and $\pvac$.  

  Since the conservation equation $\nabla_\nu T\mN =0$ for the scalar 
  field SET is automatically satisfied, we conclude that in this solution
  the function $\phi(u)$ is arbitrary, while after choosing it, the metric
  function $\alpha$ is obtained from (\ref{01s}), that is, $4\alpha_u 
  = \eps\, u\, \phi_u^2$. We see that actually a scalar field is of radiative 
  nature and represents a special case of a radiation flow considered in 
  the previous subsection: we just have 
\beq 
              \Phi = \Phi_s = \eps \e^{-2\alpha} \phi_u^2.
\eeq 
  Therefore, again, $\alpha = \alpha(u)$ under a proper choice of the $v$ 
  coordinate. As in the previous case,  the only admissible form of 
  vacuum is again a string cloud, and a Birkhoff situation automatically occurs.  

  One should also notice that canonical and phantom scalar fields can 
  form such a solution on equal grounds.

\section {Conclusion}

  We have considered some unconventional spherically symmetric geometries 
  in connection with the \BT\ and obtained the following:

\medskip\noi
  1. In the case $r = \const$, the Ricci tensor necessarily belongs to the 
  Segre types mentioned in Theorem 2, but the \BT\ holds under an
  additional physically motivated condition that the lateral pressure is functionally 
  related to the energy density. The fourth Killing vector (in addition to those 
  due to spherical symmetry) is spacelike or timelike.

  The corresponding solutions are the Bertotti-Robinson, Nariai and 
  Plebanski-Hacyan solutions and their straightforward generalizations  
  which consist in possible inclusion of Dymnikova's vacuum (D-vacuum)
  in other forms than a Maxwell field and a \cosco.

\medskip\noi
  2.  In the case $r \ne \const$ but $(\d r)^2 =0$, if we require that the geometry 
  \rf{ds-uv} is supported by some matter comoving to one of the RFs realized by
  the coordinates $t = (u+v)/2$ and $x = (u-v)/2$, then the only kind of such 
  matter is a cloud of radially aligned cosmic strings, and the metric has the form
  \rf{ds-w}.

\medskip\noi
  3. In the case $r \ne \const$ but $(\d r)^2 =0$, admitting noncomoving matter,
  there exists a much wider set of solutions with the metric \rf{ds-uv} supported by
  either an anisotropic fluid or a combination of a cosmic string cloud and null matter 
  in the form of pure radiation, in particular, a radiative massless scalar field $\phi(u)$. 

  In the case of an anisotropic fluid, an additional null Killing vector that manifests 
  a ``Birkhoff property'' of the system exists under extra (though natural) assumptions 
  that pressures and densities are functionally related by some equations of state.  
  
  With pure radiation, an additional null Killing vector exists without any extra assumptions.
  In all these cases, the structure of the Ricci tensor is more general than assumed in
  Theorem 2, so {\it these results make one more extension of the \BT}.    

  It turns out that in all such cases the total lateral pressure $p_\bot$ is equal to zero.
  
  Some kinds of matter with nonnull behavior, such as an isotropic perfect fluid or 
  scalar fields with potentials $V(\phi) \ne \const$, are incompatible with the type of
  geometry under consideration. 

\medskip\noi
  4. All solutions in the case $r \ne \const$, $(\d r)^2 =0$ contain a singularity 
  at $u =0$, as is clear from the expression for the Kretschmann invariant for 
  all of them: for the metric \rf{ds-uv} with $\alpha=\alpha(u)$ and $r \equiv u$
  it is equal to $4/u^4$.  

  The solutions obtained here, with such kinds of matter as an anisotropic fluid, 
  string clouds, scalar fields and pure radiation, seem to be new, although the 
  corresponding geometries have been previously classified from a mathematical 
  viewpoint  \cite{goenner-70, mcI-72}. 

  These solutions are not \asflat, hence they cannot describe the gravitational fields
  of isolated bodies, but one can speculate that they can represent a limiting form of
  the fields in a neighborhood of a forming horizon in the process of gravitational collapse.    
  
  The present study can be easily extended to plane and pseudospherical symmetries, 
  to an arbitrary number of dimensions, and to nonminimally coupled scalar fields.

\subsection*{Acknowledgments}

K.B. and M.S. are grateful to the colleagues from Ewha Womans University 
for kind hospitality and to Sergei Krasnikov for numerous helpful discussions. 
The work was supported in part by Ewha Womans University Research Grant 2015.
The work of K.B. and M.S. was also supported within the RUDN-University program 
5-100.

\end{document}